\documentclass[reprint,amsmath,amssymb,prl]{revtex4-1}

\usepackage{graphicx}
\usepackage{dcolumn}
\usepackage{bm}
\usepackage[english]{babel}
\usepackage[utf8x]{inputenc}
\usepackage{amsmath}
\usepackage{color}

\newcommand{\figref}[1]{Figure~\ref{#1}}

\begin{document}

\preprint{AIP/123-QED}

\title[]{Multiple-photon excitation of nitrogen vacancy center in diamond}

\author{P. Ji}
 \affiliation{Dept. of Physics and Astronomy, University of Pittsburgh, Pittsburgh PA}
\author{R. Balili}
 \altaffiliation[Corresponding author]{ email address: rbb5@calvin.edu}
 \affiliation{Dept. of Physics and Astronomy, Calvin College, Grand Rapids MI}
\author{J.Beaumariage}
 \affiliation{Dept. of Physics and Astronomy, University of Pittsburgh, Pittsburgh PA}
\author{S.Mukherjee}
 \affiliation{Dept. of Physics and Astronomy, University of Pittsburgh, Pittsburgh PA}
\author{D. Snoke}
 \affiliation{Dept. of Physics and Astronomy, University of Pittsburgh, Pittsburgh PA}
\author{G. Dutt}
 \affiliation{Dept. of Physics and Astronomy, University of Pittsburgh, Pittsburgh PA}

\date{\today}

\begin{abstract}
We report the first observation of multi-photon photoluminescence excitation (PLE) below the resonant energies of nitrogen vacancy (NV) centers in diamond. The quadratic and cubic dependence of the integrated fluorescence intensity as a function of excitation power indicate a two-photon excitation pathway for the NV$^-$ charge state and a three-photon process involved for the neutral NV$^0$ charge state respectively. Comparing the total multi-photon energy with its single-photon equivalent, the PLE spectra follows the absorption spectrum of single photon excitation. We also observed that the efficiency of photoluminescence for different charge states, as well as the decay time constant, was dependent on the excitation wavelength and power. 
\end{abstract}

\pacs{Number}
\maketitle

\section{\label{sec:level1}Introduction}
The demonstration of single photon generation \cite{Gruber12} and optical preparation and readout out of spin \cite{Jelezko04} in diamond nitrogen-vacancy centers (NV), coupled with its unique combination of features such as bio-compatibility, stability, sensitivity to magneto-electric fields and long spin coherence, has launched diamond to prominence with regards to applications for nanoscale biological sensing and carbon-based quantum technology \cite{Doherty13,Schirhagl14}. Due to an efficient absorption in the visible region, around 450 nm to 575 nm for both the NV$^0$ and NV$^-$ states \cite{Doherty13}, the electronic states of the electrons trapped in the NV defect centers are often prepared by non-resonantly exciting diamond with single photons using green light (520-540 nm). A resonant excitation of NV$^-$ zero-phonon line (ZPL) at 638 nm (1.945 eV) to photo-ionize an electron and covert the NV to its neutral charged state via an absorption of two photons followed by an Auger process has also been reported \cite{Siyushev13}. Below resonance excitation at 1064 nm, half the energy of the usual green light excitation, has been used to study nonlinear photo-physical processes \cite{Wee2007} and charge state dynamics \cite{Ji16,Hopper16}. An interesting feature observed in these works was the spin-dependent signal that is emitted when simultaneously exciting with near infra-red and visible light, a phenomenon that is not well understood. To the best of our knowledge, there has not been a systematic wavelength-dependent multi-photon photoluminescence excitation (PLE) experiment to investigate multi-photon absorption and its corresponding fluorescence. It is important to note that multiple-photon excitation is extremely advantageous for biological imaging and quantum information processing as it widely separates the signal fluorescence from excitation.\\

In this Letter, we report a first PLE measurement of NV photoluminescence using multi-photon excitation with pulsed light in the near-infrared range, 1030-1276 nm (0.97-1.20 eV). We discuss the different PL yields of NV at these excitation wavelengths. We also compare the time-resolved fluorescence decay of the NV excited states at various excitation energies and pump powers to shed light on the important features of the defect states.

\section{Experimental Setup}
\begin{figure}[ht]
\includegraphics[width=8.7cm]{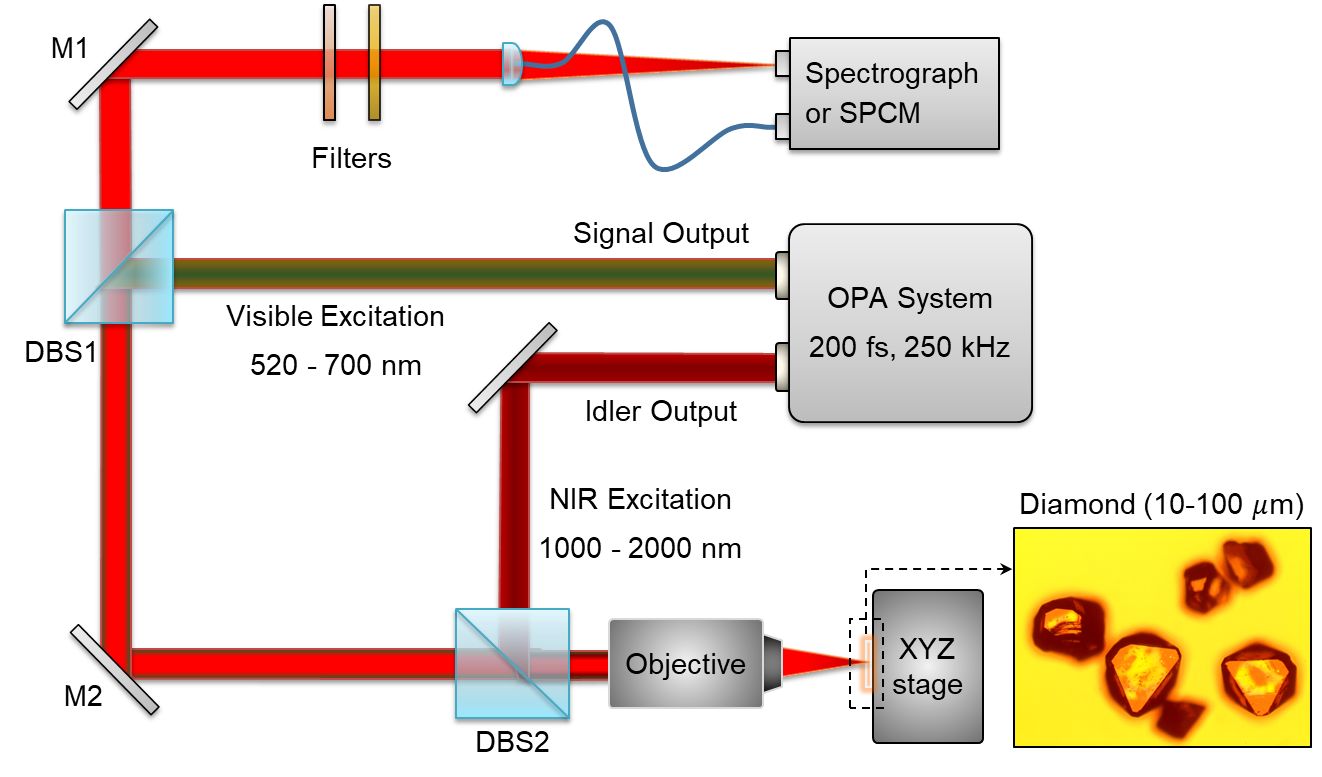}
\caption{\label{setup}An OPA system served as a tuneable light source, providing 200 ps excitation pulses at 250 kHz repetition rate. The ``signal” and ``idler” outputs of the OPA were directed onto the sample optical path using dichroic beam splitters (DBS1 and DBS2). Additional bandpass filters and longpass filters were placed along the ``signal” and “idler” paths respectively to get rid of any light leaking from the OPA’s own pump source (800 nm). Laser paths for the different wavelength regions were blocked or opened as needed in the experiments. The photoluminescence can be directed unto a spectrometer or a Single Photon Counting Module (SPCM) for lifetime measurements.}
\end{figure}
The NV ensembles investigated in these experiments are implanted within micro-diamond crystals (10-100 $\mu m$ sizes from Columbus Nanoworks) with 100 ppm nitrogen density. A laser pump, illuminating about 1 $\mu m$ spot size on these crystals, yields sufficient fluorescence signal to noise ratio despite of an inherently low multi-photon excitation efficiency. An Optical Parametric Amplifier (Coherent OPA 9400) pumped by an amplified Ti:Sapphire laser system is used to provide 200 ps pulses, about 30 nm FWHM bandwidth at 250 kHz repetition rate, in the wavelength range used for the experiments. The ``signal” output of the OPA can be tuned from 500 to 700 nm for non-resonant single-photon excitation while the ``idler” covers the excitation wavelengths between 1000 to 2000 nm. A 0.8NA objective is used to both focus the pump beam onto the sample and collect the diamond fluorescence. The samples were glued onto a translation stage and the fluorescence from the NV centers were collected by either a spectrometer for time-averaged spectroscopy or a Single Photon Counting Module (SPCM) for time correlated measurements. The trigger of the single photon counter was sourced from the 250 kHz repetition rate of the OPA pump and counts were recorded by a (Picoharp 300) dual-channel counter. A 582/75 nm bandpass filter and 650 nm longpass filter were placed along the collection path when distinguishing between the NV$^0$ and NV$^-$ photoluminescence respectively. 

\section{Results}

\subsection{Power Dependence}
\begin{figure}[hbt]
\includegraphics[width=7cm]{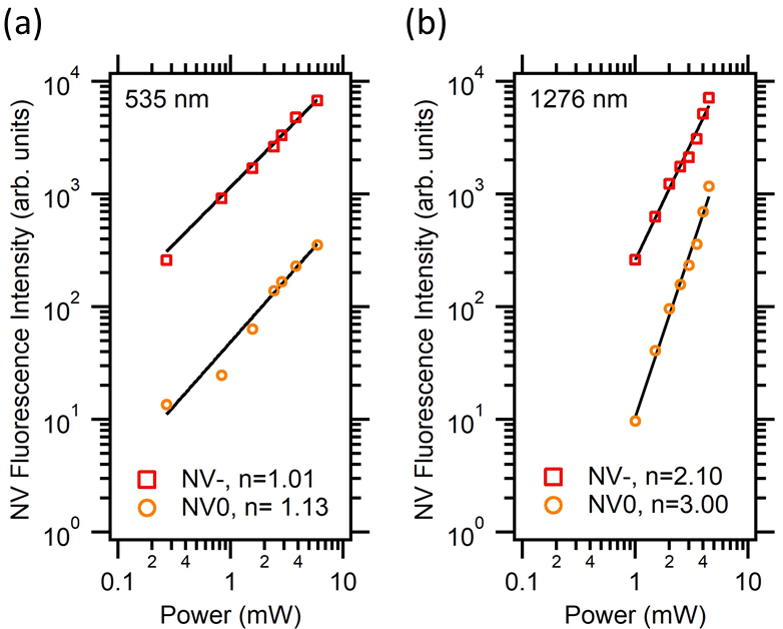}
\caption{\label{power}{Power dependence of NV$^0$ and NV$^-$ fluorescence intensities at (a) 535nm and (b) 1276nm excitation wavelengths. The solid lines are power law fits with $n$ as the power parameter.}}
\end{figure}

To identify multi-photon excitation generated PL from NV centers, we compared the power dependence of collected PL with pulsed excitation centered at 535 nm with excitation centered at 1276 nm, half the NV$^-$ ZPL energy in \figref{power}(a) and (b). The integrated counts from the spectrally resolved fluorescence for NV$^0$ (550-620 nm) and NV$^-$ (650-800 nm) as a function of power is fitted with a power law to the n$^{th}$ power. The emission intensity from the 535 nm excitation as a function of pump power yields a clear linear dependence as shown in \figref{power}(a). The NV$^-$ fluorescence follows mainly a quadratic dependence n=2.10$\pm$0.18 while the NV$^0$ fits a cubic power law n=3.00$\pm$0.34 for 1276 nm excitation. The uncertainty $\delta$n of the n$^{th}$ power is the standard deviation of several trials for the corresponding excitation wavelength. 

\begin{figure}[hbt]
\includegraphics[width=8.6cm]{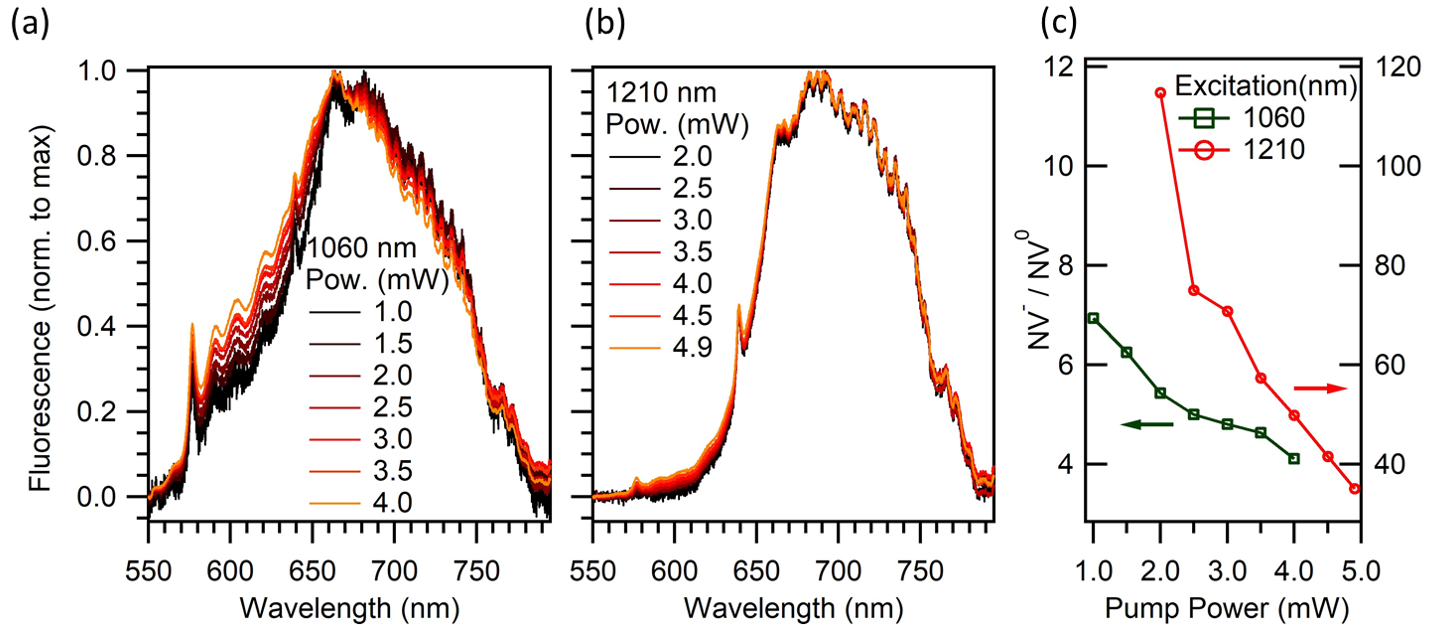}
\caption{\label{power_charge_shift}{Fluorescence of NV diamond at various pump powers for (a) 1060 nm excitation, (b) 1210 nm excitation and their corresponding NV$^-$ to NV$^0$ ratio (c).}}
\end{figure}

\figref{power_charge_shift} looks at how the near-infrared excitation affects the spectral profile of the NV system. A couple of representative fluorescence in a series of excitation powers and their corresponding NV$^-$ to NV$^0$ ratio is shown for 1060 nm and 1210 nm. At 1060 nm excitation, two photons have enough energy to excite both the NV$^-$ and NV$^0$ charge states, while past 1150 nm, e.g. 1210 nm, an additional photon is needed to reach the excited state of NV$^0$. \figref{power_charge_shift}(a),(b) shows an increase in NV$^0$ component, or equivalently a decrease in NV$^-$ to NV$^0$ ratio, with increasing excitation power. Longer excitation wavelengths past 1210 nm show more NV$^-$ than NV$^0$. It is clear that irrespective of excitation wavelength, the NV$^0$ component increases with pump power, e.g. \figref{power_charge_shift}(c), albeit at lesser rates for longer wavelengths. This indicates a higher ionization rate from NV$^-$ to NV$^0$ due to much larger illumination power used for multiple-photon excitation \cite{Aslam13}.

\subsection{Spectrum and Lifetime}

\begin{figure}[hbt]
\includegraphics[width=8.6cm]{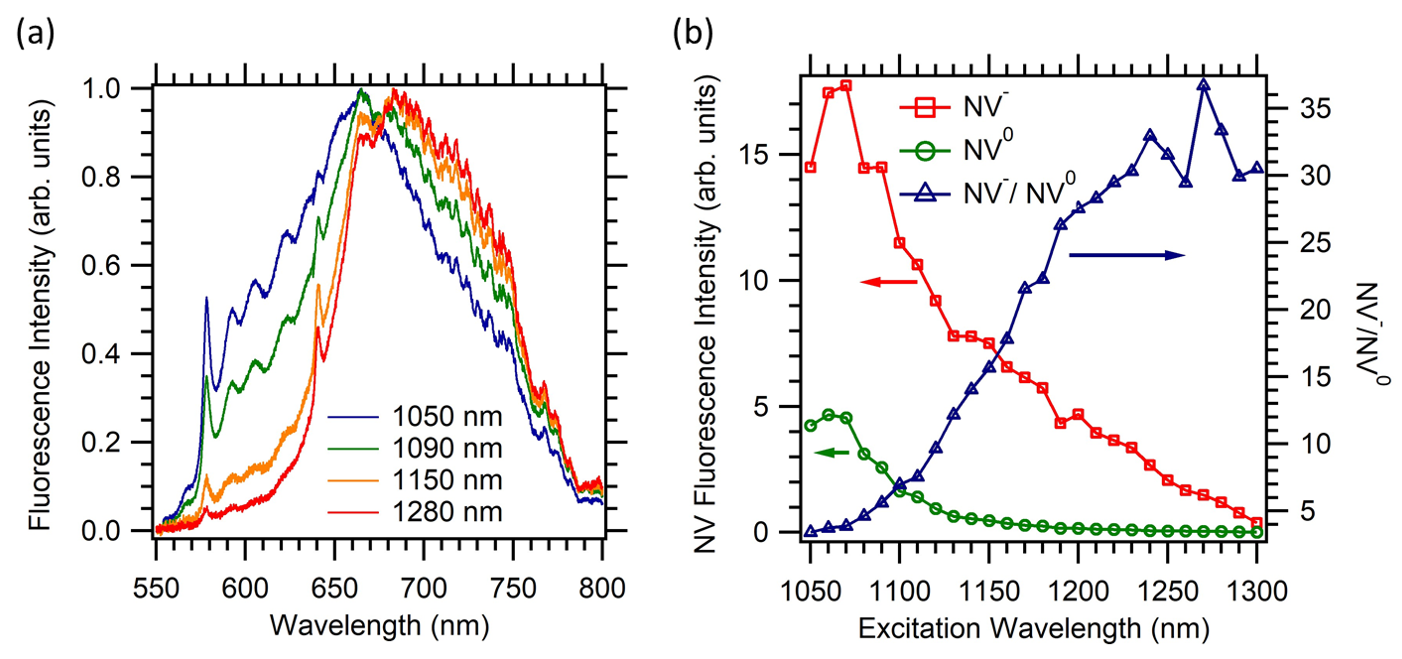}
\caption{\label{PLE}{(a) Representative PLE spectra of NV centers at 1050 nm, 1090 nm, 1150, and 1280 nm. The fluorescence spectra are normalized to maximum intensity. (b) Integrated intensity of NV0 (550-620 nm) and NV- (650-800 nm) from spectra, e.g. \figref{PLE}(a), versus excitation wavelength. The ratio between NV$^-$ and NV$^0$ intensities is also shown in the graph.}}
\end{figure}

Following the same diamond centers, we look at the wavelength dependence of the spectral profile. \figref{PLE}(a) shows representative room temperature fluorescence spectra of the NV centers at various excitation wavelengths. Our results indicate that NV centers can be excited continuously from 1030 nm to 1310 nm with varying efficiency for each of the different charged states. Using appropriate bandpass filters, we collect the integrated fluorescence intensity of the NV$^-$ charge state (750-800 nm) and the NV$^0$ charge state (550-620 nm) separately as a function of pulsed excitation wavelength. The integrated intensity for each charge state and the ratio between them is shown in \figref{PLE}(b). If an energy equivalent of two excitation photons is assumed, the emission intensity at these excitation wavelengths is consistent with the characteristic absorption cross sections or photoluminescence (PLE) spectra of NV$^-$ with conventional single-photon excitation \cite{Davies76}. For example, in our data the intensity in fluorescence emission peaks at around 1070 nm corresponding to twice the wavelength where single photon absorption peaks (535 nm) in diamond NV \cite{Beha12}. The inefficient pumping of the NV$^0$ charge state with wavelengths longer than the ZPL of the NV$^0$ (575 nm), is reflected in the decrease in fluorescence intensity as function of pump wavelength across 1150 nm towards longer wavelengths shown in \figref{PLE}(b). The fluorescence intensity of the NV$^0$ charge state ensemble goes down as well but, nonetheless, remains non-zero. A notable feature is the increase in intensity ratio (NV$^-$/NV$^0$) of the charged states toward longer wavelengths. The NV$^-$ emission and excitation is much more efficient at wavelengths beyond 1150 nm.

\begin{figure}[hbt]
\includegraphics[width=8.6cm]{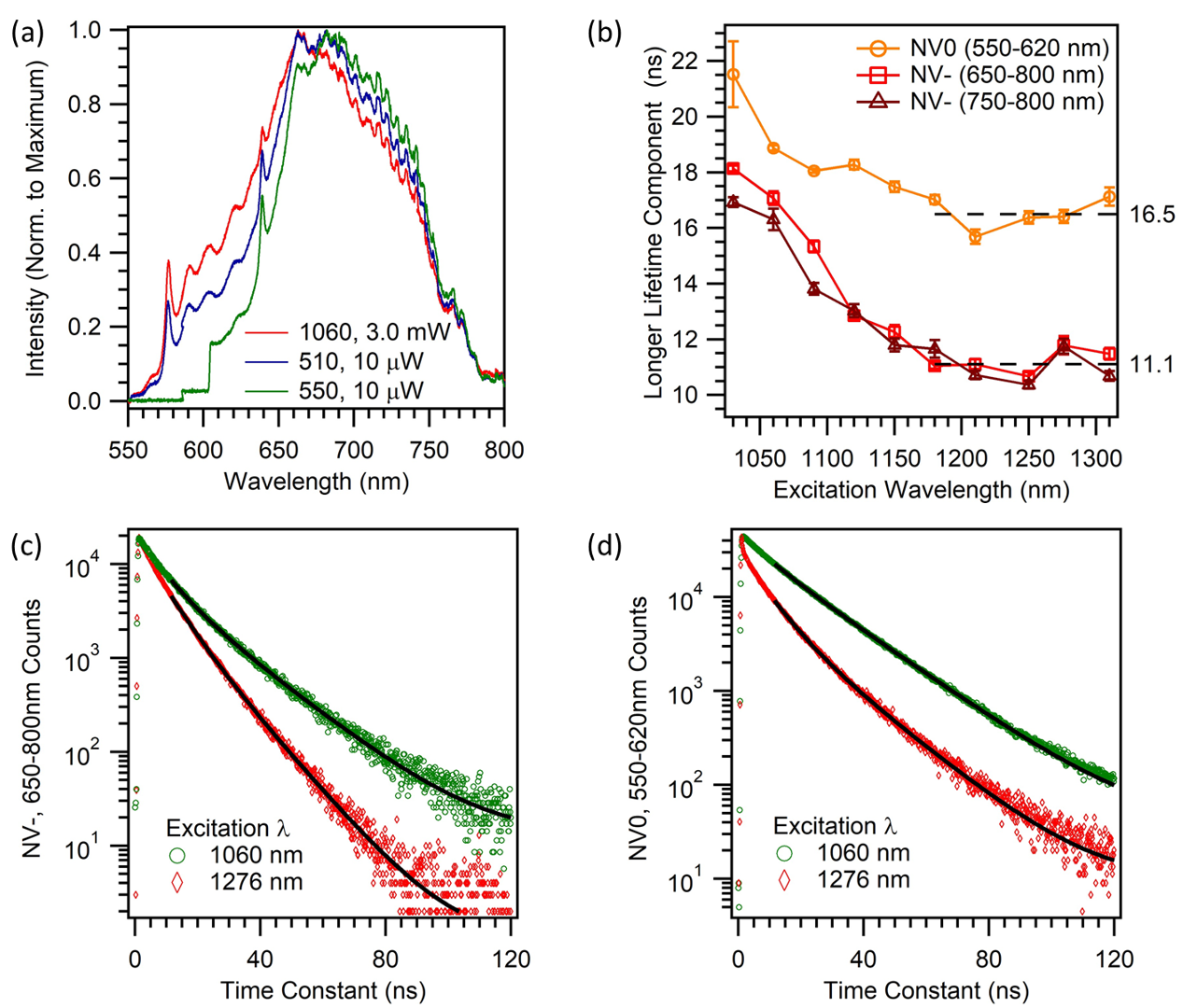}
\caption{\label{lifetime}{(a) Fluorescence comparison of single photon excitation (510 nm and 550 nm) with multi-photon excitation (1060 nm). In graph (b), the NV$^-$ fluorescence is taken at two wavelength windows for comparison. The fluorescence decay at 1 mW pump power is fitted with a bi-exponential function. The longer lifetime or slower decay component is plotted for various excitation wavelength. The shorter lifetime or faster decay component is consistent at $6.9 \pm 1.7$ ns. Representative decay plots and bi-exponential fits for (c) NV$^0$ and (d) NV$^-$ fluorescence comparing shorter, 1060 nm, versus longer, 1276 nm, excitation wavelength are also shown above. }}
\end{figure}

We also compared the NV ratio between single photon (510 nm, 550 nm) and two-photon excitation (1060 nm) as shown in \figref{lifetime}(a). Due to the laser light’s proximity to the  fluorescence region, a long pass filter with cut-off at 600 nm is placed in the collection path to filter out the laser reflections from the sample when exciting at 550 nm. The effect of the laser filter is manifested in the loss of fluorescence information below 600 nm for 550 nm excitation. Nonetheless, the reduction of NV$^0$ with respect to NV$^-$ is qualitatively apparent compared to 510 nm or 1060 nm excitation. Within the single photon excitation range, longer wavelength yields increasing NV$^-$ intensity with respect to NV$^0$ as seen when comparing 510 nm excitation with a 550 nm excitation which is consistent with the absorption spectra of the charged states \cite{Beha12}. Also, one can clearly see from \figref{lifetime}(a) that near-IR excitation contains relatively more NV$^0$ component than NV$^-$.

The decay of the fluorescence is also recorded at various wavelengths and fitted with a bi-exponential decay function. The slower decay component is plotted in \figref{lifetime}(b). The faster decay component has a lifetime of $6.9 \pm 1.7$ ns which does not vary much with excitation wavelength. For the component with slower decay, there is a noticeable change in the lifetime from shorter to longer excitation wavelength which eventually flattens beyond 1150 nm excitation. Considering an energy equivalent of two excitation photons, this downward trend in lifetime is again consistent with nanodiamond lifetime measurements with conventional single-photon excitation \cite{Storteboom:15}. The NV$^-$ fluorescence is taken at two wavelength windows for comparison, one in the 650-800 nm interval and the other at 750-800 nm interval. The consistently higher time constant for NV$^-$ 650-800 nm fluorescence, below 1150 nm excitation is indicative of some NV$^0$ fluorescence leaking into the NV$^-$ collection window. Beyond 1150 nm, the NV$^0$ is less intense minimizing the effect of leakage. We can take the time constant beyond this excitation wavelength to be the NV$^0$ lifetime 16.5 $\pm$ 0.6 ns and NV$^-$ lifetime 11.1 $\pm$ 0.4 ns, which are within the range reported by other groups.A two lifetime component can be justified by the differences in radiative lifetime between the $m_s=0,\pm1$ state or NV$^0$ and NV$^-$ PL quenching due to substitutional nitrogen defects \cite{Liaugaudas12}.The change in NV$^0$ lifetime with wavelength could be due to spin selective coupling to  the nonradiative triplet state reducing the effective lifetime \cite{Storteboom:15}. It is also possible that the variation of the longer lifetime component is an effect of other fluorescing centers or defects that are activated at certain wavelengths and subsequently quench the both the NV charge state photoluminescence.

\section{Analysis}
Experimentally, the two-photon absorption (2PA) cross section is related to the detected PL counts \cite{Xu2015}
\begin{align}
\sigma_{2} \approx \frac{Ff\tau}{ 1.28 \phi C \eta n \lambda P^{2}} 
\end{align}
where $F$ is in units of photon/s; $\phi$ is the PL collection efficiency of the experimental system; $\eta$ is the quantum yield of the defect; $C$ is the density of the defects, $\lambda$ is the excitation wavelength in vacuum; $n$ is the refractive index of the media; $P$ is the average laser power; $\tau$ is the pulse duration; $f$ is the repetition rate; and $\sigma_2$ is the  2PA cross section in units of $GM$, where $1\:GM=10^{-50}$ cm$^{4} \cdot$ s/photon.

\begin{figure}[htbp]
\centering
\includegraphics[width=5cm]{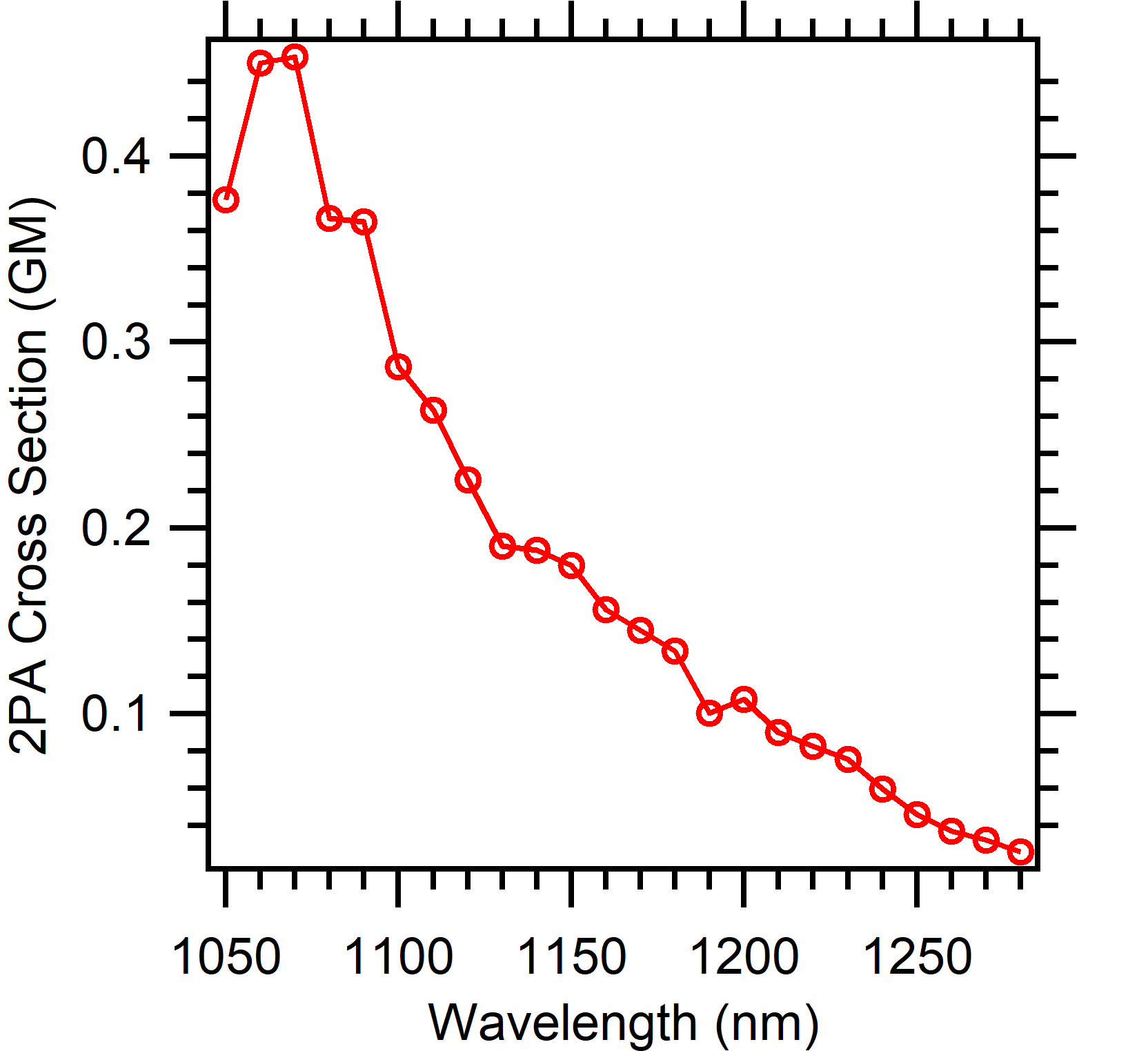}
\caption{\label{2PA_cross_section}{NV$^{-}$ 2PA absorption cross section calculated from our PLE data.}}
\end{figure}

The NV$^{-}$ 2PA cross section $\sigma_{2}(\lambda)$ from 1050nm to 1280nm  excitation can be estimated from our PLE data. Considering our laser linewidth is $\sim 30$nm, we use the measured 2PA cross section at 1064nm from reference \cite{Wee2007} as our calibration for the 1060nm excitation such that  $\sigma_{2}(1060) \approx 0.45 \pm 0.23 GM $. Assuming the experimental conditions remained fixed during the PLE measurement as the excitation wavelength is changed, then the 2PA cross section $\sigma_{2}(\lambda)$ can be simplified to
\begin{align}
\sigma_{2}(\lambda)=\frac{1060}{\lambda}\frac{F(\lambda)}{F(1060)}\sigma_{2}(1060).
\end{align}
\figref{2PA_cross_section} shows our estimate  of the 2PA cross section for NV$^{-}$ center at different excitation wavelength based on our PLE measurements. The 2PA cross section agrees with the reported one-photon absorption(1PA) by reference \cite{Beha12} at room temperature. The missing resonance at 1150nm is possibly due to the our broad laser linewidth. Low temperature experiment and narrower laser linewidth will help to explore the resonance in the future.

\begin{figure}[hbt]
\includegraphics[width=7cm]{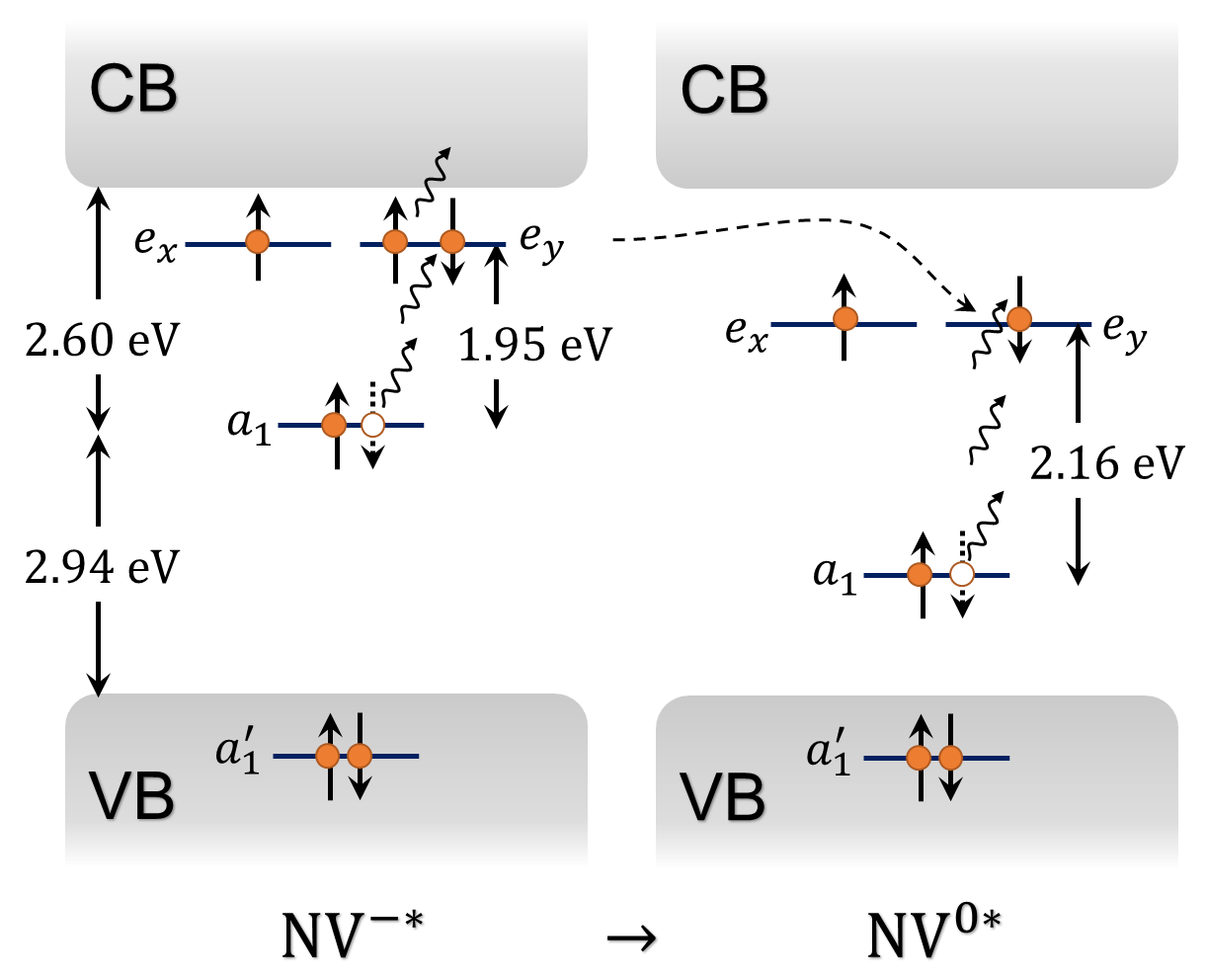}
\caption{\label{schematic}{A two-photon, three-photon excitation of NV$^-$ and NV$^0$ respectively leaving them in their excited states ($*$). An excited negatively charged state NV$^{-*}$ could absorb another photon to raise the electron out to the conduction band. If an NV$^-$ state absorbs three photons with a total energy of 2.60 eV, it is converted into the excited state of NV$^0$. A direct three-photon absorption of an NV$^0$ electron from the ground state will leave it in its excited state (NV$^{0*}$) if the total photon energy is greater than or equal to 2.16 eV.}}
\end{figure}

The schematic of the negative and neutral state shown in \figref{schematic} maps our understanding of the behavior of the system. The energy level of the states and band offsets follows reference \cite{Beha12} which is the most consistent with our results. From the schematic, we can see that an excited negatively charged state NV$^{-*}$ could absorb another photon, 0.65 eV or higher, to raise the electron out to the conduction band. Hence, if an NV$^-$ state absorbs three photons, each photon at half the energy of the NV$^-$ ZPL, it can be converted into the excited state of NV$^0$. If the total transition energy is attained, a direct three-photon absorption of an NV$^0$ electron from the ground state will also get the system to its excited state (NV$^{0*}$). 

 The power dependence of the NV$^-$ and NV$^0$ charged states with 1276 nm excitation provides evidence for a multi-photon procedure or a series of pathways for the NV electronic transitions. For the negatively charged state, a two-photon excitation or absorption is the most straightforward interpretation based on its quadratic power dependence. The cubic dependence of the NV$^0$ PL intensity on the other hand may have two possibilities. The mechanism for these is illustrated in \figref{schematic}. The first is a direct absorption of three photons 1276 nm photons to reach the excited state. Note that three times the energy of a 1276 nm photon, for a total of 2.92 eV, is more than enough to reach the 2.16 eV needed to excite the NV$^0$. The second possible explanation is an NV$^-$ to NV$^0$ conversion. One such conversion mechanism was proposed by Beha et al.\cite{Beha12} and Shiyushev et al.\cite{Siyushev13} whereby an excited NV$^-$ state absorbs another photon into the conduction band followed by an Auger process that releases enough energy to remove the defect charge. Later work, e.g. \cite{Beha12}, on photo-induced ionization dynamics did not invoke the Auger process. The NV$^-$ ionization energy was concluded, per Ref.\cite{Beha12}, to be only 2.6 eV. We point out that the three 1276 nm photons have more than enough energy required for an NV$^-$ electron to reach the conduction band from the NV$^-$ ground state. 

We also point out that the two-photon process is more efficient for exciting the NV$^0$ state than the NV$^-$ compared to excitation at the single photon equivalent energy (\figref{lifetime}(a)). This is likely due to excitation pulses which have higher peak power inducing faster charge state conversion from NV$^-$ to NV$^0$ compared with continuous-wave (CW) pumping.

The NV$^-$'s slower decay component, \figref{lifetime}(b), has apparently a shorter lifetime when excited at wavelengths beyond 1150 nm than when excited at wavelengths below 1150 nm. In part, this can be attributed to the likelihood of the NV$^0$ fluorescence significantly leaking into the NV$^-$ collection window. The leakage effect is supported by a consistently longer time constant for NV$^-$ fluorescence collected at 650-800 nm as compared to a 750-800 nm collection window. The effect can also explain the leveling of the time constant at $>$1150 nm pumping due to an inefficient two-photon excitation of NV$^0$ at these wavelengths. Less NV$^0$ leaks into the NV$^-$ collection window, unless the pump power is increased whereby 3-photon processes starts to dominate and NV$^0$ population recovers.

CW, green laser in tandem with a 650nm long pass filter is often used for NV$^-$, single-photon related experiments. However, this strategy is not effective in filtering out the NV$^0$ emission as demonstrated in \figref{lifetime}(b). A longpass filter with a much longer wavelength cutoff is needed to significantly reduce the fluorescence overlap between the two charge states. Alternatively, we propose 2P excitation with longer wavelengths, $>$1150nm, to increase the NV$^-$ ratio of the collected fluorescence.

The longer lifetime component for NV$^0$ decay for excitation wavelengths shorter than 1150 nm is indicative of multiple processes feeding into NV$^0$ population. Diamond fluorescence in a series of excitation powers and wavelengths and their corresponding NV$^-$ to NV$^0$ ratio also show that there is a significant number of carriers feeding into the NV$^0$ states at higher powers and for wavelengths shorter than 1150nm (half the energy of NV$^0$ ZPL). It is possible that this may be due to conversions from NV$^-$ to NV$^0$, excitation of defects and quenching of PL due to other defects  in the vicinity. The dependence of the lifetime with wavelength would be better understood using a streak camera which is within purview of future work.

\section{Conclusion}
We have shown that near-IR excitation of NV centers, resonant at half the energies of the NV states and wavelengths within the vicinity (1030 to 1310 nm) is allowed through a multi-photon process. Preferential population of NV$^-$ or NV$^0$ is achieved by tuning to an appropriate wavelength or excitation power. In general, longer wavelength and lower power is required to have more NV$^-$ population and vice versa for populating NV$^0$. We note that excitation at low energy is extremely useful for biomedical imaging applications since infrared light can penetrate deep into biological tissues with little scattering or damage \cite{Jacques13}. In addition, current generation of single photons are limited by spectral jumps and diffusion of emission lines caused by simultaneously exciting impurities in diamond \cite{Robledo10}. Near-IR excitation may suppress photo-ionization of surrounding impurities potentially allowing spectrally-stable, indistinguishable single photon generation required for linear optical quantum computing \cite{Knill01}. For quantum experiments using the  NV$^-$ charge state, the increase in signal-to-noise ratio as the NV$^-$ to NV$^0$ ratio increases with excitation wavelength, between 1150 to 1276 nm, is highly desirable as well. Another important application is in spin-photon entanglement experiments. NV centers can be prepared in an entangled state between a single photon and the m$_s$ = $\pm$1 electronic spin of an NV center \cite{Togan10}. Following selective resonant excitation to the A$_2$ state, the system decays into one of the spin states, simultaneously emitting a circularized polarized photon which is orthogonal to the photon emitted when an electron decays to an opposite spin state. Isolating this photon emission is significantly challenging due to the difficulty in extinguishing the resonant excitation pulse back into the detector. A near-IR excitation would widely separate the excitation and emission wavelength, thereby increasing signal contrast for the entangled photon emission.

\begin{acknowledgments}
R. Balili wants to thank the Michigan Space Grant Consortium (MSGC) seed grant and the Calvin Alumni Association (CAA) for the equipment funding and travel grants that made this collaboration possible. P. Ji was supported by the Charles E. Kaufman New Initiative grant and NSF grant PHY-1540879. G. Dutt acknowledges  DOE grant no.DE-8C-0006638 for materials, equipment, and financial support. We also acknowledge Dr. Arfaan Rampers for the helpful discussions and Columbus Nanoworks Inc. for the diamond samples.
\end{acknowledgments}

\bibliography{ref}

\end{document}